\documentclass[aps,pre,twocolumn,showpacs]{revtex4}
\usepackage{graphicx}
\usepackage{textcomp}
\usepackage{psfrag}

\begin{document}
\title{Missing physics  in  stick-slip dynamics  of  a model for 
peeling of an adhesive tape}
\author{Rumi De$^{1,\star}$}
\author{G. Ananthakrishna$^{1,2,\dag}$}
\affiliation{$^1$ Materials Research Centre, Indian Institute of Science,
Bangalore-560012, India.\\
$^2$ Centre for Condensed Matter Theory, Indian Institute  of Science,
 Bangalore-560012, India.}

\begin{abstract}
It is now known that the equations of motion for the contact point during 
peeling of an adhesive  tape mounted on a roll introduced earlier
are singular and do not  support dynamical jumps across the two stable branches of the peel 
force function.  By including the kinetic energy of  the tape in the Lagrangian, we derive   equations of motion that support stick-slip jumps as a natural consequence of the  inherent  dynamics. In the low mass  limit, these equations  reproduce solutions obtained using a  differential-algebraic algorithm  introduced for the earlier equations.
Our analysis also shows that  mass of the 
ribbon has  a strong  influence on the nature of the dynamics.  
\end{abstract}

\pacs{05.45.Pq, 62.20.Mk, 68.35.Np}

\maketitle
It is well established that there are three different modes of failure 
during peeling of an adhesive tape from a substrate \cite{MB}. At low 
applied velocities, the peeling front  keeps pace with the  
pull velocity and the failure mode is cohesive. At high  pull 
velocities, the failure is adhesive.   In between these two regimes, there is an  intermittent mode of failure corresponding to stick-slip dynamics \cite{MB,Aubrey}  accompanied by a characteristic audible noise. The stick-slip nature suggests that this regime is unstable. Indeed, the strain energy  release rate exhibits two stable  branches separated  by an unstable branch. 
Detailed   studies by Maugis and Barquins \cite{MB} and 
others \cite{HY,Ciccotti98,Ciccotti04}  show that the nature of the wave forms of the  pull force  exhibits  sinusoidal, sawtooth and highly irregular
wave patterns.  Gandur {\it et al.} \cite{Gandur97} have 
carried out a dynamical analysis of the force waveforms and  
of acoustic emission signals, and report chaotic dynamics at the 
upper end of the  pull velocities \cite{Gandur97}. Analysis of acoustic signals has also been carried out by Ciccotti {\it et al.} \cite{Ciccotti04}.

Stick-slip behavior is commonly observed in a number of driven systems
such as  the Portevin-Le Chatelier (PLC) effect ~\cite{KFA}, frictional sliding ~\cite{Perrson}, and earthquake dynamics  ~\cite{BK}. Stick-slip is characterized by the system spending most part of the time in the stuck state and a short time in the slip state. One common feature of such systems is that the force exhibits "negative flow rate characteristic" (NFRC). In fact, studies of dynamics of such systems, including that of the adhesive tape, use the
macroscopic phenomenological NFRC feature as an input, although
the unstable region is not accessible.

Maugis and Barquins  \cite{MB} were the first 
to look at  the peeling problem from a dynamical angle.  Later, 
a detailed study was carried out by Hong and Yue ~\cite{HY} on a three
 variable model originally introduced by Maugis \cite{Mreport}  using 
 an "N" shaped function that mimics the  peel force function.  
They report that the system of  equations displayed periodic and chaotic 
stick-slip solutions. However, it was later recognized that  the 
stick-slip oscillations were {\it not} obtained as a natural consequence of 
the equations of motion \cite{HY2,Ciccotti98,Rumi04}  as the jumps in the rupture speed were introduced  {\it externally}  once the velocity exceeded the limit of stability. Later, Ciccotti {\it et al.} ~\cite{Ciccotti98} interpret  stick-slip 
jumps as catastrophes with the belief that the jumps in the rupture velocity 
cannot be obtained from  these equations of motion ~\cite{Ciccotti98}.
Recently we derived these equations  starting from a Lagrangian 
and showed \cite{Rumi04} that these equations are singular and fall in the category of differential-algebraic equations (DAE) \cite{Hairer} requiring a special algorithm. Using a DAE  algorithm, we showed  that stick-slip jumps 
across the two branches arise in a pure dynamical way.  
The dynamics was also shown to be  much richer than  anticipated earlier.

However, even as the DAE algorithm offers a mathematical framework for 
obtaining solutions for these singular equations, it is difficult to 
provide any  physical interpretation for the "mass matrix"  that 
removes the singularity. {\it Thus,  a proper identification  of the missing 
physics responsible for  the absence of dynamical jumps in these equations 
still remains to be addressed}. As we shall comment later,
 this is {\it a necessary step  for  understanding the origin of 
acoustic emission  (AE) during peeling}, a problem that has remained 
 unresolved. 
A reexamination of the earlier derivation showed that the kinetic 
energy of  the stretched part of the tape was 
ignored. {\it Here, we show that the inclusion of this 
additional  kinetic energy removes the singularity thus converting them
into set of  ordinary   differential equations (ODE).
Further, stick-slip jumps emerge as a  natural consequence of the 
inherent dynamics itself.} Apart from reproducing  the DAE solutions for low 
mass limit, our analysis shows that the mass  of the tape has a
 considerable influence on the nature of the dynamics.

\begin{figure}[]
\mbox{
\includegraphics[height=2.6cm,width=4.4cm]{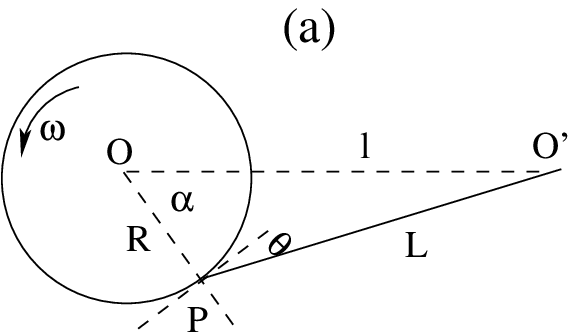}
\psfrag{v}{{\Large $v$} [m s$^{-1}$]}
\psfrag{fxxxx}{$f(v,V)$ [N]}
\includegraphics[height=3.5cm,width=4.0cm]{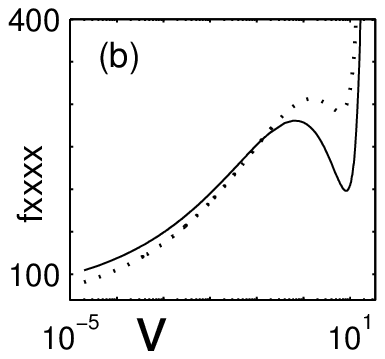}}
\caption{(a) Schematic plot of experimental setup. (b) Plots of $f(v,V)$  
for $V=1$ (solid curve), $V=4$ (dotted curve).} 
\label{tape}
\end{figure}

A schematic of the experimental  setup  is shown in 
Fig. \ref{tape}(a). An adhesive roller tape of radius $R$ is  mounted on an 
axis passing through $O$ normal to the paper and is driven at a constant 
speed $V$ by a  couple meter motor positioned at $O'$. The  pull force $F$ 
acting along line joining the contact point $P$ and  $O'$ subtends an  
angle  $\theta$ to the tangent at the point $P$. 
As the contact point $P$ moving with a local velocity $v$  
can undergo rapid bursts of velocity during rupture, the peeled length 
of the ribbon, $L = PO'$ is not fixed.
We note here that the peel force often called the force of adhesion 
denoted by $f$ is what is measured in experiments in steady state conditions.
Let the distance from the center 
of the roller tape $O$  to the motor $O'$ be $l$ and  $\alpha$ the angle 
subtended by $PO$ with the horizontal line 
$OO'$. Let $I$ be the  moment of inertia of the roller tape, $\omega$ the angular velocity, $k$ the elastic 
constant of the tape, and $u$ the elastic displacement of the tape. 
As the contact  point is not fixed, $\omega = \dot \alpha + v/R$. 
The geometry of the setup gives 
$L\;{\rm cos}\, \theta = -l\;{\rm sin}\,\alpha$ and 
$L\;{\rm sin}\,\theta = l\;{\rm cos}\,\alpha - R$.
As the peeling point $P$ moves,
the pull velocity $V$ is  the sum of  three contributions~\cite{MB}, 
i.e., $V = v + \dot u - \dot L,$ which gives
\begin{eqnarray}
v = V - \dot u + \dot L= V - \dot u - R\;{\rm cos}\,\theta\;\dot\alpha.
\label{v}
\end{eqnarray}
Then, the Lagrangian  is ${\cal L}= U_K - U_P$ with the kinetic energy 
given by $U_K={1\over2} I \omega ^2 + {1\over2} m \dot u ^2$ and the
 potential energy by  $U_P = {1\over2} k u^2$. [${1\over2} I \omega ^2 $ 
is the kinetic energy of the  roller tape and ${1\over2} m \dot u ^2$ 
ignored earlier,  arises due to the kinetic energy of the stretched part 
of the tape. $\dot u$ refers to the time derivative of $u$.  
In principle $m$ ($m=m_0 L(t)$ where $m_0$ is the mass per unit length)
depends on time through $L$.
However, for all practical purposes $m$ can be treated as constant.] 
We write the dissipation  function as ${\cal R} = \Phi(v,V)= \int f(v,V)dv$, where $f(v,V)$ assumed to be derivable from a potential function $\Phi(v,V)$, physically represents the peel force,  and is taken to depend on the pull velocity and  rupture speed as in  \cite{Rumi04}.

Using the Lagrange equations of motion, 
${d\over dt}\left({\partial{\cal L} \over {\partial\dot\alpha}}\right)-
{\partial{\cal L} \over {\partial\alpha}}+
{\partial{\cal R} \over {\partial\dot\alpha}}=0$ and
${d\over dt}\left({\partial{\cal L} \over {\partial \dot u}}\right)-
{\partial{\cal L} \over {\partial u}}+
{\partial{\cal R} \over {\partial \dot u}}=0$,
and using ($\alpha,\dot\alpha,u,\dot u$) as generalized coordinate,
we get   
\begin{eqnarray}
 \ddot \alpha &=& - {\dot v \over R}  + {R \over I} {{\rm cos}\, \theta \over (1- {\rm cos}\, \theta)} { f(v,V)},\label{ddalpha} \\
m\ddot u &=& {1 \over (1 -{\rm cos }\, \theta)} [  \, f(v, V) 
-k \,u (1- {\rm cos}\, \theta ) ]. \label{ddu}
\end{eqnarray}
\noindent
Note that the right hand side of Eq. (\ref{ddu}) is the 
algebraic constraint in Eq. (10) of Ref. \cite{Rumi04} or Eq. 5(d)  of Ref. \cite{HY}.
These equations in their present form are still not suitable for further 
analysis as  they have to satisfy the constraint Eq. (\ref{v}). 
In the spirit of classical mechanics of systems with constraints (see 
Ref. \cite{ECG}), we derive the equation for the acceleration  variable $\dot v$ 
in the constraint equation  by differentiating
Eq. (\ref{v}) and using Eqs. (\ref{ddalpha}), and (\ref{ddu}). This gives  equations of
 motion for $\alpha, \omega, u$ and $v$ 
\begin{eqnarray}
\dot\alpha &=& \omega - {v/R}, \label{flow1}
\\
\dot \omega &=&  { R \over I}{{\rm cos}\, \theta \over (1- {\rm cos}\, \theta)}  f(v,V), \label{flow2}
\\
\dot u &=& V - v - R\;{\rm cos}\, \theta \, \dot \alpha, \label{flow3}
\\
\dot v &=& - \ddot u + R\;{\rm sin}\, \theta \, \dot \theta \, \dot \alpha  - R\;{\rm cos}\, \theta \,\ddot \alpha. 
\label{flow4a} 
\end{eqnarray}
Using Eqs. (\ref{ddalpha}), (\ref{ddu}) in Eq. (\ref{flow4a}), we get
\begin{eqnarray}
\nonumber
 \dot v & = & {1 \over (1- {\rm cos}\, \theta)} \Big[{ku \over m} -  {f(v,V) \over m(1- {\rm cos}\, \theta)}  
        - {(R cos \theta) ^2 f(v,V) \over I (1- {\rm cos}\, \theta)}  \nonumber \\
         &+ & {R\over L} \dot \alpha ^2 (l  cos \alpha - R  (cos \theta)^2)\Big].
       \label{flow4}
\end{eqnarray}

\begin{figure}
\mbox{
\psfrag{v}{{\Large $v$} [m s$^{-1}$]}
\psfrag{F}{$F$ [N]} 
\includegraphics[height=3.5cm,width=4.5cm]{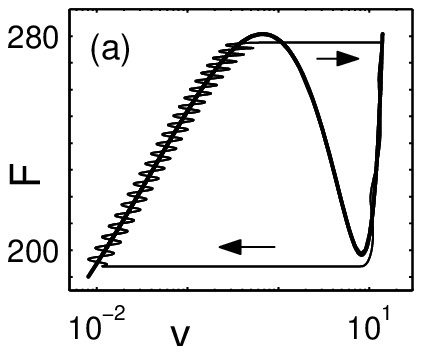}
\psfrag{v}{{\Large $v$} [m s$^{-1}$]}
\includegraphics[height=3.5cm,width=4.0cm]{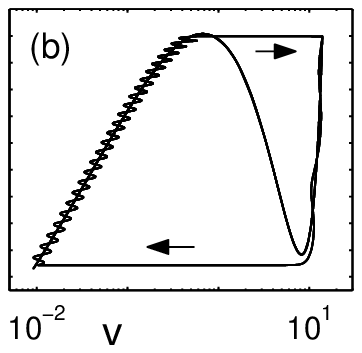}}
\caption{(a), (b) The $v-F$ phase plots for $I = 10^{-5}, V = 1$ 
corresponding to the DAE and ODE solutions  for $m=10^{-4}$. Solid 
line shows $f(v,1)$.} 
\label{vFsmallI}
\end{figure}

We retain the dynamization scheme introduced earlier \cite{Rumi04} adopted 
from the PLC effect \cite{Chihab} wherein   the difference between 
the maximum and minimum of $f(v,V)$ is assumed to decrease with increasing $V$. The  parameterized  form of $f(v,V)$ used here is given by
\begin{eqnarray}
f(v,V) &=& 402v^{0.34} + 171v^{0.16} + 68e^{(v/7.7)}- 2V^{1.5} \nonumber\\
&& -(415 - 45V^{0.4} - 0.35V^{2.15})v^{0.5}. \label{fvV}
\end{eqnarray}
Equation (\ref{fvV})  mimics the general trend of the experimental 
peel force function and  is essentially the same 
as used earlier \cite{Rumi04} except 
that it  accommodates larger excursions of  
trajectories arising from the introduction of the additional time scale [Fig. \ref{tape}(b)]. 
The fixed point of Eqs. (\ref{flow1}, \ref{flow2}, \ref{flow3}, \ref{flow4}) given by $\alpha=0, \omega=V/R, u=f(V,V)/k$ and $v=V $ becomes  unstable when $V$ is such that  $f'(V,V)<0$ leading to a Hopf bifurcation.

We have solved Eqs. (\ref{flow1}), (\ref{flow2}), (\ref{flow3}), and (\ref{flow4}) by adaptive step size stiff differential equations 
solver (MATLAB package) and  studied the dynamics  over a 
wide range of  values of $I$ (kg m$^{2}$), 
$V$ (m s$^{-1}$) for $m$ (kg) ranging from $10^{-4}$ to 0.1 (keeping 
$ k = 1000$ N m$^{-1}$,  $R = 0.1$ m, and $l = 1$ m).
(Henceforth, we suppress the units for the sake of brevity.) 
Here, we present results (obtained  after 
discarding the initial transients) for a few representative values of the 
$m$ when ($I,V$) are at low and high values.  We  show that 
for low  $m$, we essentially recover the DAE solutions reported earlier 
\cite{Rumi04}. Further we show that mass of the tape has a strong 
influence on the nature of the dynamics.

\begin{figure}
\psfrag{v}{\Large $v$}
\psfrag{F}{\large $F$}
\includegraphics[height=3.6cm,width=7.0cm]{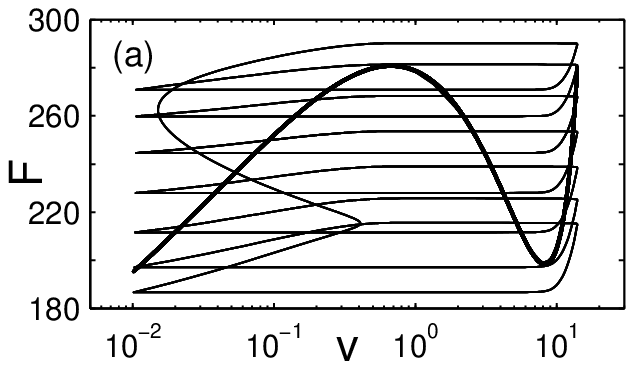}
\psfrag{v}{\Large $v$}
\psfrag{F}{\large $F$}
\includegraphics[height=3.5cm,width=7cm]{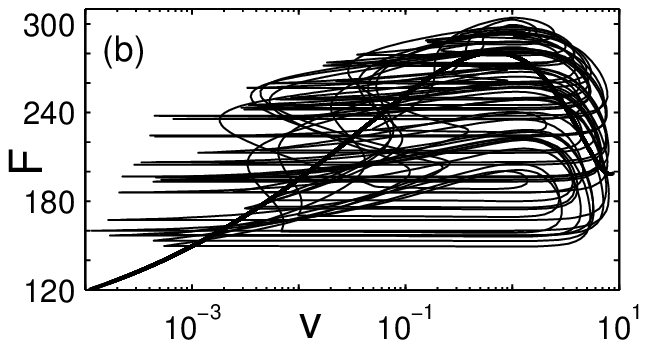}
\hbox{
\psfrag{v}{\Large $v$}
\psfrag{t}{\large $t$}
\includegraphics[height=3.5cm,width=4.0cm]{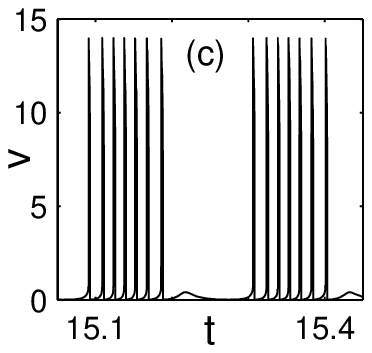}
\psfrag{t}{\large $t$}
\includegraphics[height=3.4cm,width=4.0cm]{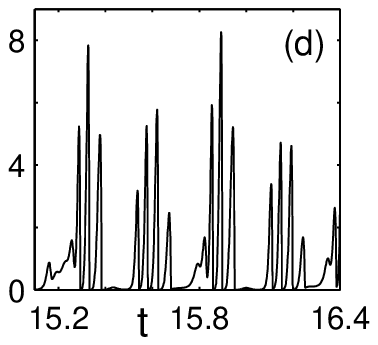}}
\includegraphics[height=3.0cm,width=7cm]{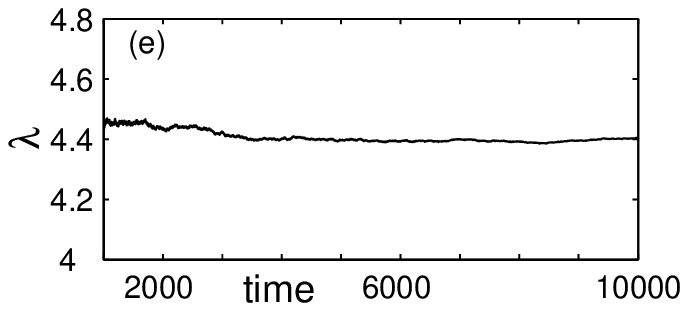}
\caption{(a), (b) Phase plots of $v-F$ obtained for $m =10^{-4}$ and 0.1 
for  $I = 10^{-2}, V = 1$. Solid line shows $f(v,1)$. (c), (d) Plots 
of $v(t)$ for $m =10^{-4}$ and 0.1. (e) The largest Lyapunov 
exponent, $\lambda$ for $m =0.1, I = 10^{-2}, V = 1$. (Units of $v$, $V$ 
are in m s$^{-1}$, $F$ in N, $I$ in kg m$^2$, $t$ in s and $\lambda$ in s$^{-1}$.)}
\label{vFlargeI}
\end{figure}

A rough idea of the nature of the dynamics can be obtained  
by  comparing the frequency $ \Omega_u =(k/m)^{1/2}$ associated 
with $u$ with $\Omega_{\alpha} =(Rf/I)^{1/2}$ corresponding 
to $\alpha$. Since $f$ (in N) is limited to 180-280, the range 
of $\Omega_{\alpha}$ (s$^{-1}$) is 
 1342-1673  for small $I$ ($\sim 10^{-5}$) decreasing to 42-53 for 
large $I$ ($0.01$). In comparison,  $\Omega_u $ (s$^{-1}$) is 3162 for $m = 10^{-4}$ 
 decreasing to 100 for $m =0.1$. Thus,  one expects 
the nature of solutions to be influenced with increasing $m$ for 
fixed $I$ which also  depends of $V$ \cite{Rumi04}.  

\begin{figure}
\psfrag{v}{\Large $v$}
\psfrag{t}{\large $t$}
\includegraphics[height=3.5cm,width=7.5cm]{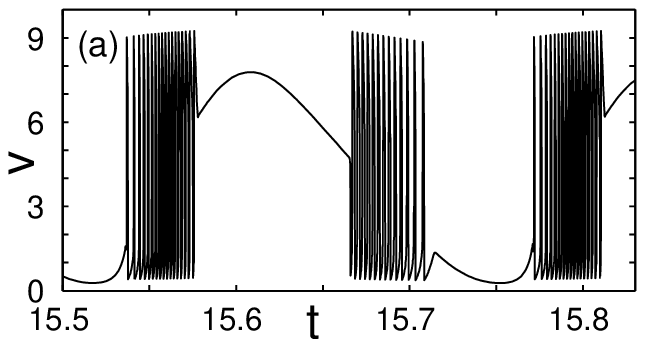}
\psfrag{v}{\Large $v$}
\psfrag{t}{\large $t$}
\includegraphics[height=3.5cm,width=7.5cm]{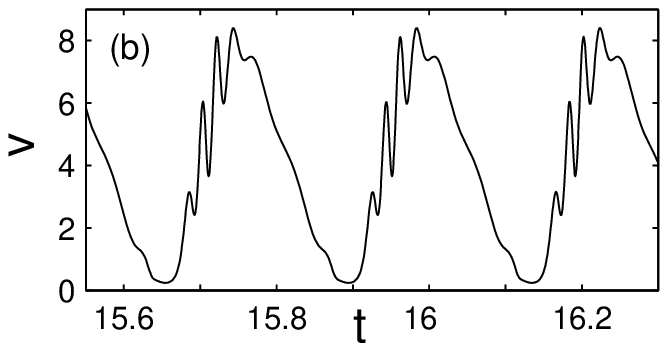}
\mbox{
\psfrag{v}{\Large $v$}
\psfrag{F}{\large $F$}
\includegraphics[height=3.6cm,width=4.5cm]{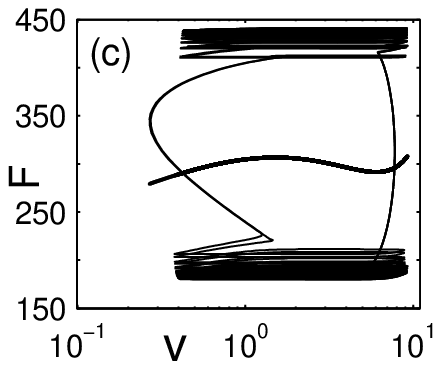}
\psfrag{v}{\Large $v$}
\includegraphics[height=3.4cm,width=4.0cm]{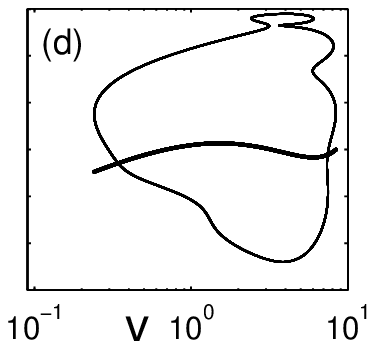}}
\caption{(a), (b) Plot of $v(t)$ for $m =10^{-4}$ and 0.1 
for $I = 10^{-2}, V = 4$. (c), (d) The corresponding phase space 
trajectories. $f(v, V)$ is shown by a bold line. (Units of $v$, $V$ are 
in m s$^{-1}$, $F$ in N, $I$ in kg m$^2$ and $t$ in s.)}
\label{vFlargeV}
\end{figure}

Here, we present our numerical results. Consider the low mass limit, i.e.,
low kinetic energy of the tape. Then the RHS of Eq. (\ref{ddu})
 is small, and turns 
$f(v, V) -k \,u (1- {\rm cos}\, \theta ) \approx f(v, V) - F \, (1- {\rm cos}\, \theta )=0$ (which is the algebraic constraint that makes the equations 
singular). Thus, one  expects that the DAE  solutions 
are reproduced for small
 $m$ which we have verified for the entire  range of values of $I$
 and $V$ studied previously  \cite{Rumi04}. (For numerical calculations,
 we have used  $m= 10^{-4}$ as the low mass limit.) As an example 
Figs. \ref{vFsmallI}(a) and \ref{vFsmallI}(b) show  the phase plots 
in the $v-F$ plane obtained using the DAE algorithm and 
ODE  Eqs. (\ref{flow1}), (\ref{flow2}), (\ref{flow3}), and (\ref{flow4}) for $m =10^{-4}$ respectively keeping  $I = 10^{-5}, V = 1$. It is evident that the DAE solution is  similar to  the ODE solution.

Much more complex dynamics  emerges as a result of a competition  between 
this additional time scale and other time scales present in the system.  
Consider the  results for $m = 10^{-4} $ and $0.1$, for   $I=10^{-2}$ and  
$V=1$. The small $m$ plots are provided  for the sake of comparison as they
 essentially correspond to the DAE solution.  Consider the phase plots 
in $v-F$ plane shown in Figs. \ref{vFlargeI}(a) for $m = 10^{-4}$ and Fig. \ref{vFlargeI}(b) for $m = 0.1$. It is 
clear that the influence of increasing $m$  is considerable. In particular, 
note that the sharp changes  in $F$  in the $v-F$ plot [Fig. \ref{vFlargeI}(a)] at the upper end of $v$  for small $m$ are  rendered smooth for 
large $m$ case [Fig. \ref{vFlargeI}(b)].   Indeed, the effect of the additional time scale
 due to  finite mass of the tape  is  also  evident in the plots of $v(t)$ 
for the low and  high mass cases shown in Figs. \ref{vFlargeI}(c) and \ref{vFlargeI}(d) respectively. 
Finally, it is clear that the phase plot [Fig. \ref{vFlargeI}(b)] fills the space and is 
suggestive of  chaotic dynamics. The chaotic nature  can be ascertained  by 
 calculating the  Lyapunov spectrum.  Using the $QR$ decomposition 
method \cite{Eck}, we have calculated the Lyapunov spectrum and find
 a large  positive exponent with a value $\sim 4.4 s^{-1}$ [Fig. \ref{vFlargeI}(e)].

\begin{figure}[!t]
\psfrag{m}{\large $m$}
\psfrag{I}{\large $I$}
\includegraphics[height=2.8cm,width=8.8cm]{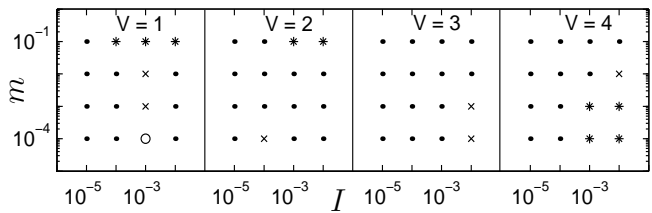}
\caption{Phase diagram in the $I - m$ plane for various 
values of $V$. Periodic $\bullet$, long 
periodic $\times$, chaotic (marginal) $\circ$ and chaotic $\ast$. (Unit 
of $I$ is in kg m$^2$ and $m$ in kg.)}  
\label{Lyp}
\end{figure}

Increasing $m$ does not always increase the level of complexity of the 
solutions. As an example, Figs. \ref{vFlargeV}(a) and \ref{vFlargeV}(b) show plots of $v(t)$ 
for $m =10^{-4}$ and $0.1$ respectively for $I= 10^{-2}$ for $V =4$. 
While the solution for small $m$ is similar 
to that of  DAE which exhibits several sharp 
spikes in velocity [Fig. \ref{vFlargeV}(a)], for 
large mass ($m=0.1$), $v(t)$ is surprisingly simple and is 
periodic [Fig. \ref{vFlargeV}(b)]. Indeed, this is better seen in  the phase plots $v-F$ 
for $m=10^{-4}$ and 0.1 shown in Figs. \ref{vFlargeV}(c) and \ref{vFlargeV}(d) respectively. In contrast 
to the low mass $v-F$ plot which is chaotic (see similar DAE solution 
in Fig. 5(c) in Ref. \cite{Rumi04}),  a simple limit cycle emerges for  $m =0.1$. 
As the nature of the dynamics can vary from a simple limit 
cycle to a chaotic attractor as the three parameters are varied, these 
results  can be summarized as phase-diagrams in the $I-m$ plane 
for different values of pull velocities, $V$, 
as shown in Fig. (\ref{Lyp}).  Apart from the chaotic state ($\ast$) seen for a few values of the parameters, for most values, the system is
 periodic ($\bullet$) and a few other values, the attractor is  long 
periodic ($\times$). We also find a (marginally) chaotic 
attractor ($\circ$) for $V=1,m= 10^{-4}, I = 10^{-3}$ for 
which the positive Lyapunov exponent is $\sim ~0.03$ s$^{-1}$ (which is much beyond the error in computation).

In summary, we have demonstrated that the missing  time scale arising from 
the kinetic energy of the stretched part of the tape plays an important role 
in the peeling dynamics of the  adhesive tape. As the inclusion of this term 
lifts the singularity  in the equations of motion hitherto considered, 
stick-slip jumps across the two resistive branches emerge as a consequence 
of the inherent dynamics.  Further, our study shows that 
the mass of the tape has a strong influence on the nature 
of the dynamics. For low pull velocities, 
and high $I$, the complexity  increases, i.e., trajectories that are not chaotic for low mass become chaotic with increasing $m$. In contrast, for high $V$, the trajectories that are chaotic for low $m$ are rendered nonchaotic with increase of $m$.

Apart from resolving the central issue,  the inclusion of  kinetic energy 
of the tape  provides a mechanism  for converting the potential energy 
stored in  the stretched tape into kinetic energy and hence provides a 
basis for explaining acoustic emission during peeling. This involves 
first extending the model to include the spatial degrees of freedom 
of the peel front and including an additional (rupture) velocity dependent 
dissipation function to mimic the AE energy dissipated (along the lines 
in \cite{RajeevRumi}). The extended model also helps to analyse the contact 
line dynamics of the peeling front which in itself has not been addressed so 
far. Preliminary results  \cite{Rumi05} show that the energy dissipated 
occurs in bursts similar to the nature of  AE signals seen in 
experiments \cite{Ciccotti04}.
Here, it is worthwhile to comment on the dynamization of the friction law.
The physical origin of this can be attributed to the viscoelastic nature of
the fluid which in turn implies frequency dependent elastic constant. 
Thus as higher pull speed allows lesser time 
for internal relaxation to be complete, the viscoelastic fluid behaves 
much like an elastic solid.
Clearly, a rigorous derivation of the peel
force function from microscopic considerations that includes the
effect of the viscoelastic glue at the contact point is needed to
understand the dynamics appropriately.

RD wishes to thank Rangeet Bhattacharyya for many useful
discussions. This work is financially supported 
by DST-SP/S2K-26/98, India.\\
\thanks{rumi}{$^{\star}$Electronic mail: rumi@mrc.iisc.ernet.in}\\
\thanks{garani}{$^{\dag}$Electronic mail: garani@mrc.iisc.ernet.in}

\end{document}